\documentclass[preprint,review,11pt]{elsarticle}
\usepackage{amssymb,amsmath}
\usepackage{graphicx}
\usepackage{epstopdf}
\usepackage{epsfig}
\usepackage{color}
\usepackage{lineno}
\usepackage{subcaption}
\usepackage{siunitx}
\usepackage{natbib}
\usepackage{lineno}
\usepackage{lscape}
\usepackage{multirow}

\usepackage[version=3]{mhchem}

\usepackage[figuresright]{rotating}

\usepackage[colorlinks=true,
            linkcolor=red,
            urlcolor=blue,
            citecolor=gray]{hyperref}

\journal{Atmospheric Pollution Research}

\begin{document}
%

\begin{frontmatter}



\title{Wet scavenging process of particulate matter ($PM_{10}$): A multivariate complex network approach}

\author[ks,ua]{Thomas Plocoste\corref{cor1}}
\ead{thomas.plocoste@karusphere.com}

\author[cg]{Rafael Carmona-Cabezas}
\ead{f12carcr@uco.es}

\author[cg]{Eduardo Guti{\'e}rrez de Rav{\'e}}
\ead{eduardo@uco.es}

\author[cg]{Francisco Jos{\'e} Jiménez-Hornero}
\ead{fjhornero@uco.es}

\cortext[cor1]{Corresponding author}

\address[ks]{Department of Research in Geoscience, KaruSphère SASU, Abymes 97139, Guadeloupe (F.W.I.), France}

\address[ua]{Univ Antilles, LaRGE Laboratoire de Recherche en Géosciences et Energies (EA 4539), F-97100 Pointe-à-Pitre, France}

\address[cg]{Complex Geometry, Patterns and Scaling in Natural and Human Phenomena (GEPENA) Research Group, University of Cordoba, Gregor Mendel Building (3rd ﬂoor), Campus Rabanales, 14071, Cordoba, Spain}


\begin{abstract}

	This paper reports the results of research on $PM_{10}$ wet scavenging by rainfall using a new multilayer complex networks called Multiplex Visibility Graphs (MVG). To the best of our knowledge, this work is the first to assess $PM_{10}$ wet deposition using multivariate time series according to African dust seasonality. We considered 11 years of daily $PM_{10}$ and rainfall data from the Guadeloupe archipelago. To analyse the impact of rainfall on $PM_{10}$ behaviour, two MVG parameters were computed: the average edge overlap ($\omega$) and the interlayer mutual information ($I_{PM_{10}Rainfall}$). On the 1-d scale, the $\omega$ results showed that the wet scavenging process was higher during the second half of the year when the high dust season and the rainy season are juxtaposed. This highlights a greater correlation between the microscopic structure of the signal, and the impact of rainfall on $PM_{10}$ concentrations is more significant when the atmosphere is loaded with dust. The joint probability computed between the $PM_{10}$ and rainfall nodes confirmed this trend. The $I_{PM_{10}Rainfall}$ results indicated a correlation between $PM_{10}$ and rainfall structures throughout the year. Furthermore, $I_{PM_{10}Rainfall}$ values were higher during the transition periods between winter and summer (and vice versa). Our study showed that MVG is a powerful technique for investigating the relationship between at least two nonlinear time series using a multivariate time series.

\end{abstract}

\end{frontmatter}
%

\section{Introduction}	
\label{intro}

	In geoscience, precipitation is a key component of the water cycle \citep{schneider2014} and of atmospheric circulation \citep{kidd2011}. In recent decades, the removal of atmospheric Particulate Matter ($PM$) by falling precipitation has greatly interested the scientific community \citep{gonzalez2012, ouyang2015, wu2018}. This phenomenon, which can occur through liquid (rain) and solid (snow) forms of precipitation, is called “wet deposition” \citep{kim2012, singh2016}. Numerous studies have shown that wet scavenging of $PM$ by rainfall is one of the primary precipitation processes for wet deposition \citep{laouali2012, tiwari2012, yoo2014, singh2016, olszowski2017, wu2018, mcclintock2019}. Raindrops falling through the air column, bump into and collect air particles. Raindrops approach the particles, apply a force via the air as a medium, and change trajectory \citep{sonwani2019}. The collision between the raindrops and the $PM$ is conditioned by size and relative location \citep{olszowski2017}. The two primary wet scavenging mechanisms related to rainfall are rainout (in-cloud scavenging) and washout (below-cloud scavenging) \citep{dallarosa2005, tombette2009, sonwani2019}. Studies have shown that wet scavenging can remove 30\% of the aerosols from the troposphere \citep{murakami1983, schumann1989}.
	
	Over the past decades, two types of $PM$ have received special attention due to their health impact: fine particles (particulate diameter $<$ 2.5 $\mu$m, $PM_{2.5}$) and coarse particles (particles with diameters between 2.5 and 10 $\mu$m, $PM_{10-2.5}$) \citep{bayraktar2010, plocoste2019a}. Epidemiological studies reveal that short- and long-term exposure to high concentrations of $PM_{2.5}$ and $PM_{10}$ can cause human health problems \citep{weinmayr2010,atkinson2014, lu2015}. The authors focused on $PM_{10}$, which also strongly impacts climate \citep{plocoste2020, plocoste2020c}.
	
	In the Caribbean, air quality is frequently degraded by African dust \citep{euphrasie2020}. Dust haze episodes primarily occur during summer \citep{petit2005, prospero2014}. Many studies present the processes that allow the transport of dust over the Atlantic Ocean \citep{perry1997, prospero1999, prospero2003, engelstaedter2006, kumar2014, euphrasie2020}. As the wet scavenging of $PM_{10}$ is a standard indicator of air quality in a given area, the aim of this study was to investigate the wet scavenging process of $PM_{10}$  by rainfall in the Caribbean Basin. Additionally, we aim to determine whether there is a link between wet scavenging efficiency and African dust seasonality.
	
	A newly developed method termed Multiplex Visibility Graph (MVG) was used to perform this study \citep{lacasa2015}. The methodology is based on a previous technique called Visibility Graph (VG), first introduced by \cite{lacasa2008}. The main idea is to transform a time series into a complex network, which can later be analysed, and to preserve some of the original information. Most of the variants of this method focus on analysis of a single time series \citep{luque2009, lan2015, carmona2019a, iacovacci2019} and have been applied to applications related to univariate time series \citep{mali2018, carmona2019b, plocoste2021a}. However, owing to their stochastic properties, atmospheric processes are frequently related to numerous degrees of freedom; that is, their behaviour is governed by a multivariate time series. To overcome this drawback, the MVG technique applies the visibility approach to examine nonlinear multivariate time series \citep{lacasa2015}. After transforming the time series into complex networks, the results were used to build a multi-layered structure that could be analysed. Owing to recent advances in the theory of multilayer networks \citep{bianconi2013, kivela2014, battiston2014, lacasa2015}, additional information can be retrieved from the original multivariate time series. To the best of our knowledge, no study has yet investigated $PM_{10}$ wet scavenging using a multivariate time series. Here, 11 years of daily $PM_{10}$ and rainfall data from the Guadeloupe archipelago were analysed.

\section{Site and data collection}	
\label{sitedata}

	The Guadeloupe archipelago ($16.25^\circ$N $-61.58^\circ$W) is a French overseas region located in the central Caribbean Basin \citep{plocoste2019b}. The small territory ($\sim$1,800 $km^2$; 390,250 inhabitants) has an insular tropical climate with meteorological characteristics that vary by location due to microclimates  \citep{bertin2013}. According to the K{\"o}ppen-Geiger climate classification \citep{peel2007}, Guadeloupe is in the ‘‘Af (tropical rainforest)’’ category.
	
\begin{figure}[h!]
\centering
\includegraphics[scale=0.8]{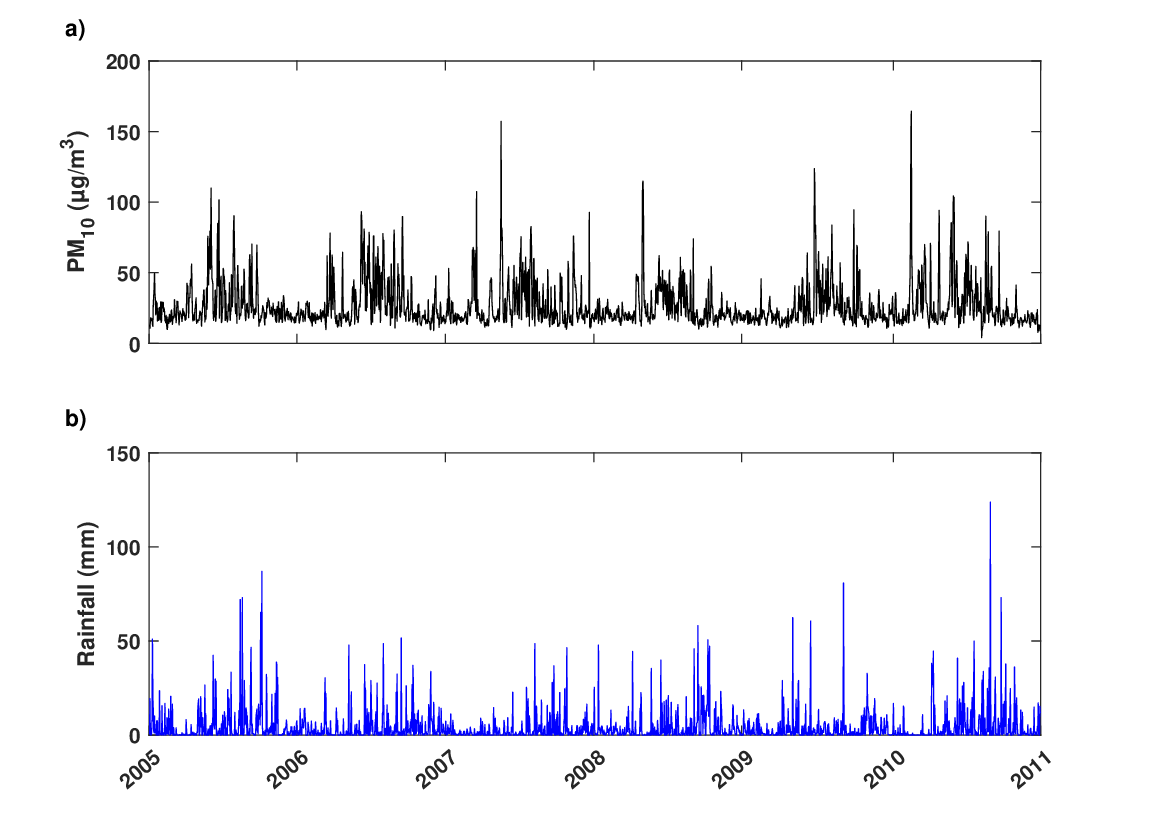}
\caption{\label{signal} Illustration of simultaneous measurement sequences between (a) daily average $PM_{10}$ concentrations and (b) the daily sum of rainfall between 2005 and 2011.}
\end{figure}

	For this study, time series of Particulate Matter ($PM_{10}$) and rainfall were used. Hourly $PM_{10}$ data were provided by Gwad'Air Agency (http://www.gwadair.fr/), which manages the Guadeloupe air quality network. $PM_{10}$ concentrations were measured using the Thermo Scientific Tapered Element Oscillating Microbalance (TEOM) models 1400ab and 1400-FDMS. From 2005 to 2017, the air quality network-principally located at the centre of the island-has only one $PM_{10}$ sensor at Pointe-\`a-Pitre ($16.2422^\circ$N $61.5414^\circ$W) from 2005 to 2012 and at Baie-Mahault ($16.2561^\circ$N $61.5903^\circ$W) since 2015. Because of the proximity between the air quality stations ($\sim$5.5 $km$), $PM_{10}$ measurements were performed under the same environmental conditions. Rainfall measurements were made by M\'et\'eo France at the international airport of P\^{o}le Cara\"{i}bes at Abymes ($16.2630^\circ$N $61.5147^\circ$W) using a Precis-Mecanique 3070. As with the $PM_{10}$ time series, the M\'et\'eo France observations are an hourly rainfall time series. Both measurements were made in the insular continental regime \citep{plocoste2018, plocoste2020d}. To assess the possible wet scavenging phenomenon over an entire day, hourly $PM_{10}$ data were converted into daily average values, whereas rainfall data were converted into daily average and daily sum values. By computing the Pearson correlation coefficient between the daily average $PM_{10}$ and the daily rainfall data (sum then average), the same result was obtained ($R$= -0.14). Many studies demonstrate the cumulative effect of rainfall on atmospheric processes \citep{winstanley1973, johnson2000}. In addition, to account for hours with and without rainfall (0 mm) in a day, the authors favoured the sum over the average for the stochastic analysis. Thus, 11 years of simultaneous measurements between the daily average $PM_{10}$ and the daily sum of rainfall were available for this study (a total of 3,849 points per time series). Figure \ref{signal} shows the sequence of the analysed time series. A slight lag appears to exist between the groups of peaks.

\section{Theoretical framework}	
\label{method}

\subsection{Visibility graphs}
\label{vgmetho}

	A graph is a mathematical object composed of a set of vertices (or nodes) which are connected by a set of lines or edges. A relatively recent tool called Visibility Graph (VG) allows the transformation of two-dimensional sets of points into graphs or networks \citep{lacasa2008}. VG has great applicability for time series analysis and produces networks that inherit many of the properties of the original time series \citep{lacasa2010}.
	
	The points in the time series correspond to the nodes in the graph. The edges of the graph (which connect nodes) are selected by checking which pairs of points meet the visibility criterion, which is as follows: two points from the time series ($t_a$, $y_a$) and ($t_b$, $y_b$) are connected only if any other point ($t_c$, $y_c$) located between them ($t_a < t_c < t_b$) fulfils the following relationship \citep{lacasa2008}:

\begin{equation}
y_{c} < y_{a} + (y_{b}-y_{a}) \frac{t_{c}-t_{a}}{t_{b}-t_{a}}
\label{VG}
\end{equation}

To construct the graph, this algorithm was applied to every pair of points in the signal. Two consecutive nodes are always connected, because there are no intermediate points.

	A graph is commonly expressed via its adjacency matrix, whose rows store the information of each node. If an element $a_{ij}$ is equal to 1, nodes $i$ and $j$ are connected, the opposite is true if $a_{ij}$ is equal to 0. In the case of a time series with N points, the resulting VG is represented by an N$\times$N  adjacency matrix, which has special properties that facilitate computation; the adjacency matrix is symmetric ($a_{ij} = a_{ji}$) and hollow ($a_{ii} = 0$); additionally, all the nearest neighbours are visible to one other ($a_{ij} = 1$ for $j = i \pm 1  $). In general, the adjacency matrix has the following form \citep{carmona2019b}:

\begin{equation}
V= 
 \begin{pmatrix}
  0 & 1 & \cdots & a_{1,N} \\
  1 & 0 & 1 & \vdots \\
  \vdots  & 1  & \ddots & 1 \\
  a_{N,1} & \cdots & 1 & 0
 \end{pmatrix}
\label{Mat}
\end{equation}

\subsection{Degree centrality}
\label{degtho}

	Degree is the most commonly used of the principal properties that can be studied from a graph and one of the centrality parameters used to measure the importance of different nodes in the graph with relation to the rest of them, using different criteria \citep{latora2017}. The degree of a node ($k_i$) measures (in an undirected graph) the number of nodes that are reciprocally connected to a given node. By considering the adjacency matrix, the degree can be computed as $k_i = \sum_{j} a_{ij}$. 
	
	Once the degree of every point is computed, a degree probability distribution $P(k)$ can be obtained for the graph (here, VG). $P(k)$ accounts for the probability of having each value of degree in the graph. To obtain information on the nature of the series, the degree distribution is analysed \citep{lacasa2008, mali2018, pierini2012}. If the right tail of the degree distribution (for high values of degree) can be fitted by a power law such as $P(k) \propto  k^{-\gamma}$, the time series has a fractal nature \citep{lacasa2008}. This part of the distribution is related to the hubs (nodes with the highest degrees), which are, by definition, rare in a graph. The exponent in the power law is the coefficient, which is related to the Hurst exponent in series related to Brownian motion \citep{lacasa2009}.

\subsection{Multiplex visibility graph}
\label{methodMVG}

\begin{landscape}
\begin{figure}[h!]
\centering
\includegraphics[scale=1.25]{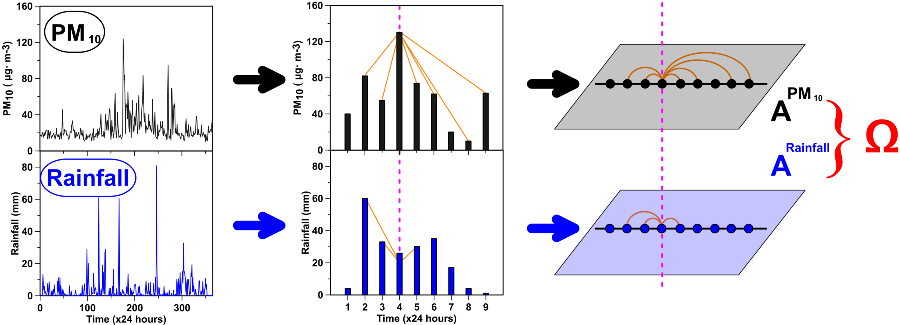}
\caption{\label{MVG} $PM_{10}$ and rainfall time series (left) are converted into complex networks with the VG algorithm (centre), which is described by an adjacency matrix ($A^{PM_{10}}$ and $A^{Rainfall)}$). Then, both are combined to design a two-layered MVG, called $\Omega$ (right image).}
\end{figure}
\end{landscape}

	Another application of VGs in the context of multivariate analysis, is the use of multi-layered networks. This methodology was recently introduced as \textit{Multiplex Visibility Graph (MVG)} \citep{lacasa2015}. The main idea behind MVG is to build each of the layers M with VGs from the different variables of the study. Therefore, as VG is represented by its adjacency matrix, so MVG is identified by a vector of adjacency matrices $\Omega=\lbrace A^{[1]}, A^{[2]},..., A^{[M]} \rbrace$. In the last expression, $A^{[\alpha]}$  corresponds to the VG adjacency matrix of the VG in the $\alpha$-dimension (or layer in the multiplex), which comes from the $\alpha$ variable of the multivariate time series (see Figure \ref{MVG}, where $PM_{10}$ and rainfall sample time series are transformed for illustrative purposes). 
	
	After construction, MVG is analysed to obtain information regarding the system of the time series. The two measures used for such purposes \citep{nicosia2015} and chosen for this work are \textit{Average Edge Overlap} ($\omega$) and \textit{Interlayer Mutual Information} ($I_{\alpha,\beta}$). $\omega$ averages the number of layers on which a given edge between a pair of nodes can be found. $I_{\alpha,\beta}$ measure the correlations between the degree distributions of the given layers $\alpha$ and $\beta$. In this study, the layers correspond to daily $PM_{10}$ concentrations and total rainfall.
	
	The computation is relatively straightforward after the MVG and degree distributions of each layer are obtained. Equation \ref{omega} shows the formula to compute the $\omega$ of a given MVG \citep{lacasa2015}:
	
\begin{equation}
\omega=\frac{\sum_{i} \sum_{j>i} \sum_{\alpha}{a_{ij}^{[\alpha]}}}{M \sum_{i} \sum_{j>i}\left(1-\delta_{0,\sum_{\alpha}{a_{ij}^{[\alpha]}}}\right)}
\label{omega}
\end{equation}
\\
	All quantities were previously defined in the text; $\delta_{0,\sum_{\alpha}{a_{ij}^{[\alpha]}}}$ corresponds to a Kronecker delta, which is 1 when $\sum_{\alpha}{a_{ij}^{[\alpha]}}$ is null, and otherwise 0. The maximum value of $\omega=1$ indicates that all the layers and, therefore, the time series are identical. Conversely, the minimum possible value of $\omega=1/M$ indicates that every edge in the MVG can be found only in a singular layer. Overall, this quantity provides an idea of the expected number of layers on which an edge can be found. In addition, a high $\omega$ value indicates a high correlation in the microscopic structure of the signal \citep{lacasa2015}.
	
	Additionally, $I_{\alpha,\beta}$ is defined in Equation \ref{interlayer} \citep{lacasa2015}:
	
\begin{equation}
I_{\alpha,\beta}=\sum_{k^{[\alpha]}} \sum_{k^{[\beta]}} P(k^{[\alpha]},k^{[\beta]})log\frac{P(k^{[\alpha]},k^{[\beta]})}{P(k^{[\alpha]})P(k^{[\beta]})}
\label{interlayer}
\end{equation}
\\
	where $P(k^{[\alpha]},k^{[\beta]})$ is the joint probability of having a degree of $k^{[\alpha]}$ in layers $\alpha$ and $k^{[\beta]}$ in layer $\beta$ that can be obtained using the following formula:
	
\begin{equation}
P(k^{[\alpha]},k^{[\beta]})=\frac{N_{k^{[\alpha]},k^{[\beta]}}}{N}
\label{joint}
\end{equation}

where $N_{k^{[\alpha]},k^{[\beta]}}$ is the number of nodes with a degree of $k^{[\alpha]}$ in layer $\alpha$ and $k^{[\beta]}$ in layer $\beta$; $N_{k^{[\alpha]},k^{[\beta]}}$ is divided by N, which is the total number of nodes or points in the time series.

\section{Results and Discussion}
\label{results}

\subsection{Preliminary analysis}
\label{resPreAn}

\begin{figure}[h!]
\centering
\includegraphics[width=0.80\textwidth]{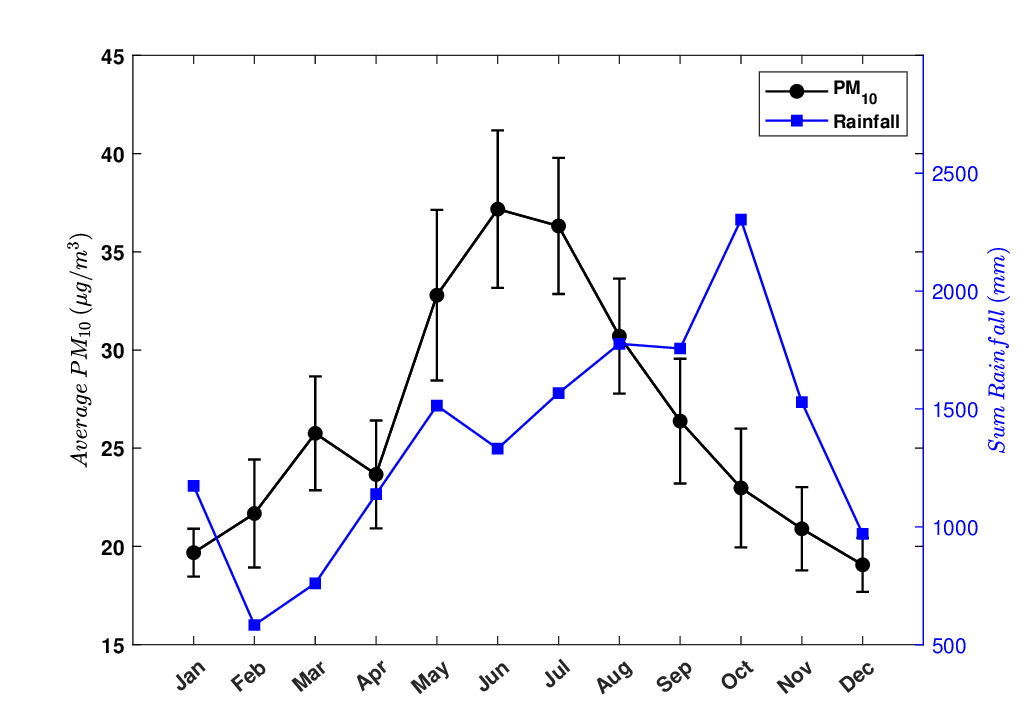}
\caption{\label{moymonth} Monthly conditional average of $PM_{10}$ and rainfall time series over the 11-y period; here, a representative year for one decade. The whiskers depict the standard deviations.}
\end{figure}

	We conducted a preliminary analysis to investigate $PM_{10}$ and rainfall seasonality throughout the year. Figure \ref{moymonth} illustrates the monthly average $PM_{10}$ concentrations and the monthly summation of rainfall data over 11-y period. Seasonality is observed in both curves, with a high dust season from May to September \citep{plocoste2020} and a rainy season from July to November \citep{bertin2013}. \cite{van2020} observed the same behaviour for both parameters in Barbados. The Inter-Tropical Convergence Zone (ITCZ) dynamics throughout the year play a key role in these seasonal behaviours. In summer, the activation of dust sources from the Saharan and Sahelian deserts coupled with the upward northward movement of the ITCZ ($10-20^\circ$N) \citep{moulin1997, adams2012, euphrasie2020} allows the transport of dust plumes from the African coast to the Caribbean area \citep{petit2005, prospero2014, euphrasie2021}. According to a statistical study lasting over a decade \citep{plocoste2020a}, average $PM_{10}$ and kurtosis are 1.5 times higher and 5.5 times lower during the high dust season, respectively, due to the recurrence of dust plumes compared with the low dust season. From October to April, $PM_{10}$ concentrations are primarily related to marine aerosols \citep{clergue2015, rastelli2017}, because of the insular context of the Guadeloupe archipelago. These aerosols are advected by trade winds which blow continuously from east to west across the Atlantic Ocean \citep{plocoste2014, plocoste2020d}. Consequently, the contribution of marine aerosols to $PM_{10}$ concentrations remains constant throughout the year. Figure \ref{moymonth} shows that the standard deviations exhibit their lowest values from October to April. Thus, marine aerosols are one of the primary constituents of the $PM_{10}$ background atmosphere \citep{plocoste2021b}. The ITCZ movement toward the north generates precipitation carried by trade winds during the boreal summer (the rainy season) \citep{giannini2000, munoz2008}. During the boreal winter (mid-January to March), the ITCZ awakens the Azores anticyclone due to its southerly movement, which reduces cloud generation (the dry season) \citep{bertin2013}.
	
\subsection{Degree distribution}
\label{resdegree}

\subsubsection{Overall analysis}
\label{resover}

\begin{figure}[h!]
\centering
\includegraphics[scale=0.70]{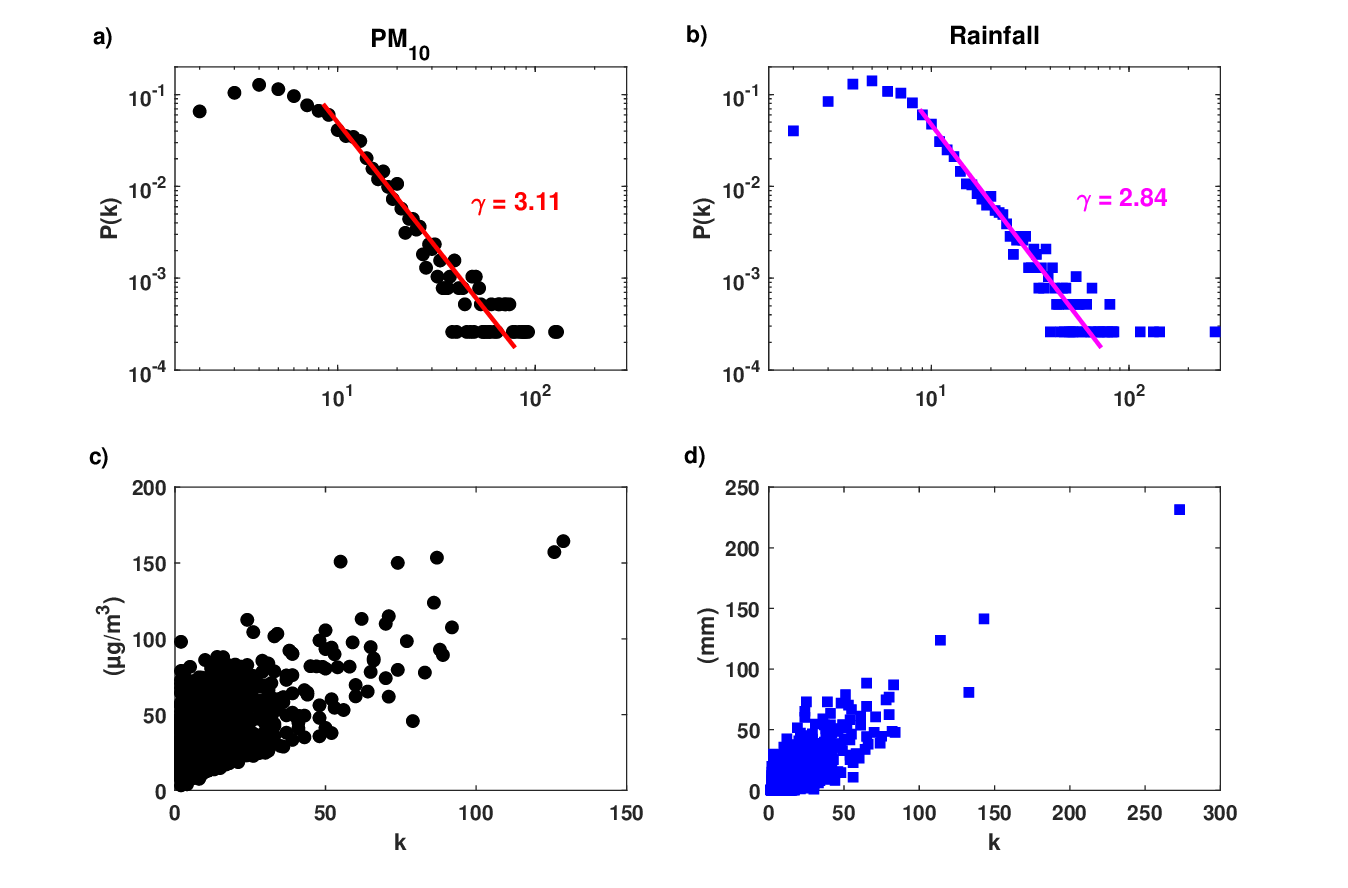}
\caption{\label{resDeg} At the top, degree distribution of the visibility graph for (a) $PM_{10}$ and (b) rainfall in a log-log plot for all the data. A the bottom, the relationship between the time series values and their degrees in (c) $PM_{10}$ and (d) rainfall.}
\end{figure}

	Before performing a profound analysis of the impact of rainfall on $PM_{10}$ concentrations in the MVG frame, both time series were analysed separately in the VG frame. The classical first approach is to study the degree distribution $P(k)$ of each time series. Figure \ref{resDeg}(a) and \ref{resDeg}(b) show the degree distributions obtained for the $PM_{10}$ and rainfall time series over the 11-y period, respectively. Both plots highlight the fractal nature of the time series. The tail region of $P(k)$ in the log-log plot can be fitted by a power law, such as $P(k) \propto  k^{-\gamma}$, where $\gamma_{PM_{10}}$ and $\gamma_{rainfall}$ equal 3.11, and 2.84, respectively. In the literature, the fractal nature of the $PM_{10}$ \citep{dong2017, nikolopoulos2019, plocoste2021a} and rainfall \citep{olsson1993, breslin1999, maskey2016} data has been observed.   

	To assess the behaviour of the highest degree (so-called hubs), the time series values versus their degrees (v-k plot) were analysed for each parameter as introduced by \cite{pierini2012}. Figure \ref{resDeg}(c) and \ref{resDeg}(d) illustrate the v-k plot for the $PM_{10}$ and rainfall data, respectively. In both cases, the hubs were related to the highest values of each time series. \cite{carmona2019b} found the same tendency for hubs of a tropospheric ozone time series in Cadiz, Spain. The value of precipitation has an almost linear relationship to the degree of rainfall. Thus, the degree of the rainfall nodes can be used to identify both high and low rainfall values. In addition, the $PM_{10}$ dot distribution appears more heterogeneous, because of the wide annual variability of African dust haze \citep{plocoste2017, plocoste2020c}.

\subsubsection{Monthly analysis}
\label{resmont}

\begin{figure}[h!]
\centering
\includegraphics[scale=0.55]{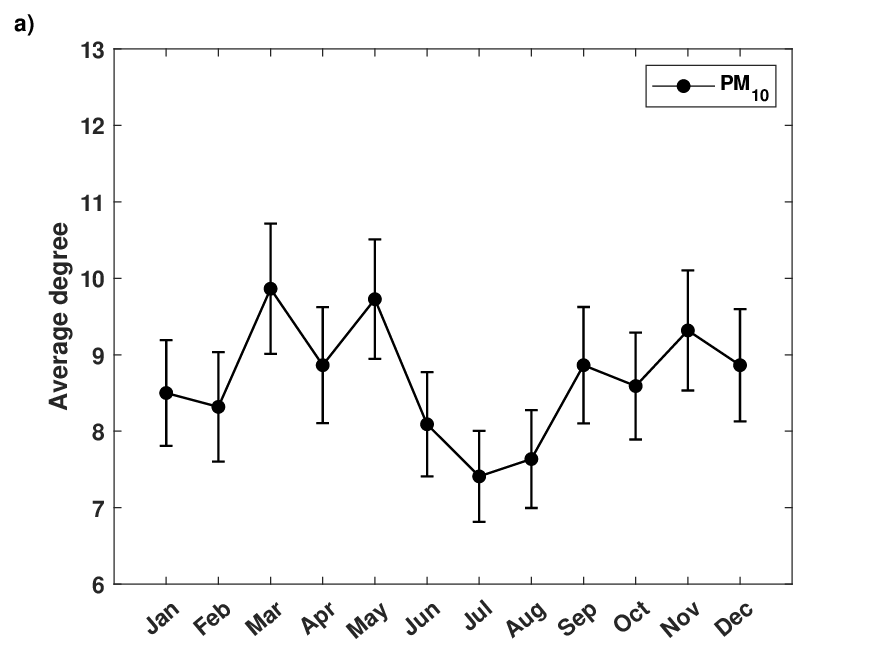}
\includegraphics[scale=0.55]{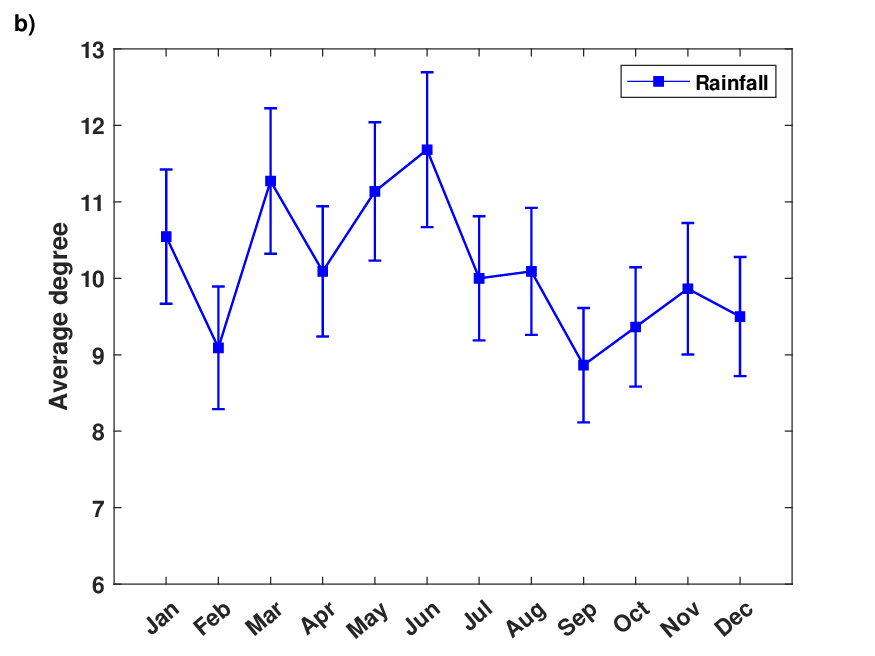}
\caption{\label{resmontDeg} Computed average degree and standard deviation from the degree distribution of each month over the 11-y period for (a) $PM_{10}$ and (b) rainfall. Each monthly value is the average of the computed 11-y values.}
\end{figure}

	We use the first centrality measure (degree centrality) to study the importance of the node for $PM_{10}$ and rainfall throughout the year \citep{carmona2019b}. Figure \ref{resmontDeg}(a) and \ref{resmontDeg}(b) highlight the monthly behaviour of the average degree and standard deviation from the degree distribution of the $PM_{10}$ and rainfall time series. A trend merged in both curves. The decay of $PM_{10}$ hubs begins at the onset of the high dust season (May-September) \citep{plocoste2020, plocoste2021a}; the decay for rainfall hubs begins at the onset of the hurricane season (June–October) \citep{tartaglione2003, dunion2011} in the Caribbean Basin. Figure \ref{moymonth} shows that the monthly behaviour of $PM_{10}$ and rainfall over the period of a decade confirms this trend, increases in $PM_{10}$ and rainfall begin in May and June, respectively. The above-mentioned results show the impact of seasonality on node distribution and highlight that the VG frame is sensitive to time-series behaviour.

\subsection{Multiplex visibility graph}
\label{resdegree}

	After performing the analysis of the $PM_{10}$ and rainfall univariate time series in the VG frame (i.e., transformation of the time series into a complex network), both complex networks were combined to design a two-layered multivariate network. Here, we investigate the wet scavenging process of $PM_{10}$ by rainfall. The authors focused on two approaches, which demonstrate the abundance of single edges across layers (average edge overlap) and the presence of interlayer correlations of the node degrees (interlayer mutual information) \citep{lacasa2015, nicosia2015}. The first approach measures the overall coherence in the multivariate time series, and the second evaluates structural correlation.

\subsubsection{Average edge overlap analysis}
\label{resomega}

\begin{figure}[h!]
\centering
\includegraphics[scale=0.65]{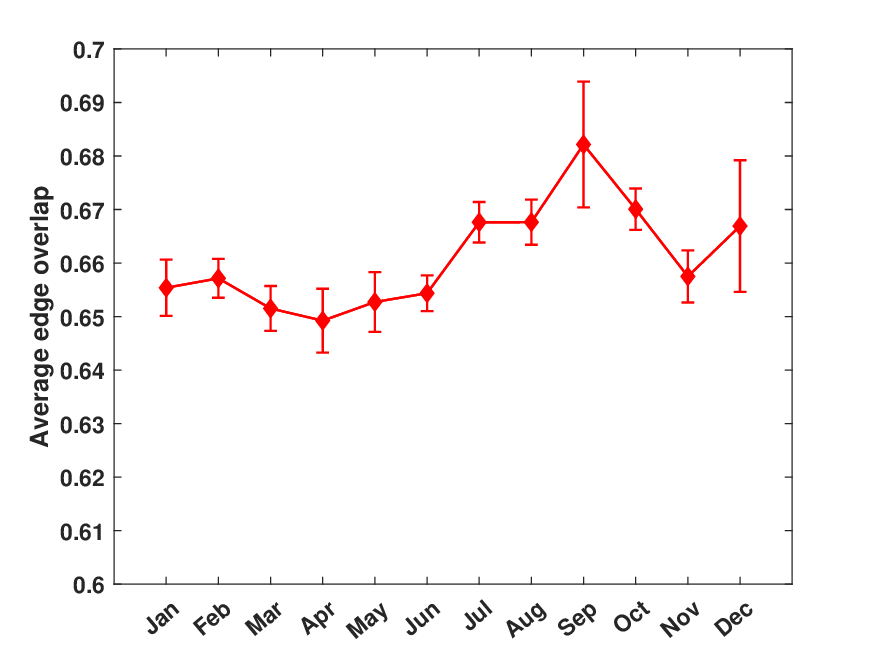}
\caption{\label{resmontomega} 11-y monthly average edge overlap values ($\omega$) and their standard deviations.}
\end{figure}

	Figure \ref{resmontomega} illustrates the monthly average edge overlap values ($\omega$) and their standard deviations over the 11-y period. $\omega > 1/M$ and $\omega < 1$; thus, the two layers are different, and edges can be found in both layers. These two criteria prove that there is an interaction between $PM_{10}$ and rainfall in the MVG frame. $\omega$ is a sensitive parameter with small  variations (increases and decreases) \citep{lacasa2015}. Here, $\omega$ was almost constant from January to June. $\omega$ values are higher from July to December and peaked in September, indicating a greater correlation in the microscopic structure of the signal \citep{lacasa2015, carmona2020} that represents interaction on the order of 1-d, which is the minimum time resolution. Therefore, at the 1-d scale, the wet scavenging process of $PM_{10}$ by rainfall is more significant during the last six months of the year. A 20-y precipitation study in the Luquillo Mountains of Puerto Rico \cite{mcclintock2019} also found a summer maximum in wet dust deposition. Physically, these findings make sense, as the summer corresponds to the high dust \citep{plocoste2020} and rainy \citep{bertin2013} seasons in the Caribbean Basin. The atmosphere is loaded with dust, and the impact of rainfall on $PM_{10}$ is greater. \cite{tiwari2012} observed that low $PM_{10}$ concentrations in New Delhi occurred during the monsoon (August-September) season due to the washout phenomenon. A 10-y study of air pollutants ($PM_{10}$, $CO$, $NO_2$, $SO_2$, and $O_3$) and precipitation over South Korea highlighted that $PM_{10}$ is most effectively scavenged by summertime rainfall due to its particulate nature \citep{yoo2014}. Because of data availability, determining which wet scavenging process (rainout or washout) is more efficient is difficult \citep{pillai2002, tombette2009, bayraktar2010}. According to \cite{sonwani2019}, the level of aerosols in and under clouds at the time of precipitation is crucial, as it determines whether both phenomena occur simultaneously.
		
	During the high dust season, the wet scavenging phenomenon naturally reduces $PM_{10}$ concentrations in the atmosphere. Due to the impact on respiratory and cardiovascular diseases, diminishing $PM_{10}$ concentrations after dust outbreaks is crucial \citep{gurung2017, zhang2017, momtazan2019, feng2019}. African Easterly  Waves (AEWs) \citep{prospero1981, plocoste2021b}, which precede and follow the dust plumes, are the principal generator of precipitation during the high dust period \citep{dominguez2020} and regulate $PM_{10}$ concentrations in the Caribbean area.

\subsubsection{Interlayer mutual information analysis}
\label{resinterlayer}

	The relationship between $PM_{10}$ concentrations and rainfall can also be determined by studying the node distribution in the MVG layers. Equation \ref{interlayer} shows that the joint probability between the $PM_{10}$ and rainfall nodes ($P(k^{[PM_{10}]},k^{[Rainfall]})$) is a building block of the interlayer mutual information ($I_{PM_{10},Rainfall}$). Thus, we first computed the joint probability before performing the interlayer mutual information analysis. 
	
\begin{figure}[h!]
\centering
\includegraphics[scale=0.55]{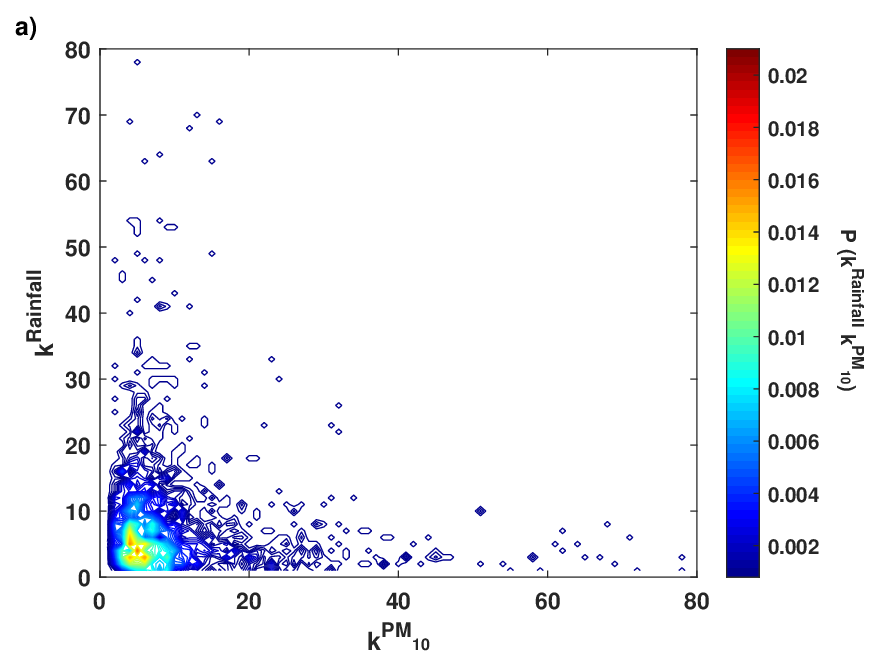}
\includegraphics[scale=0.55]{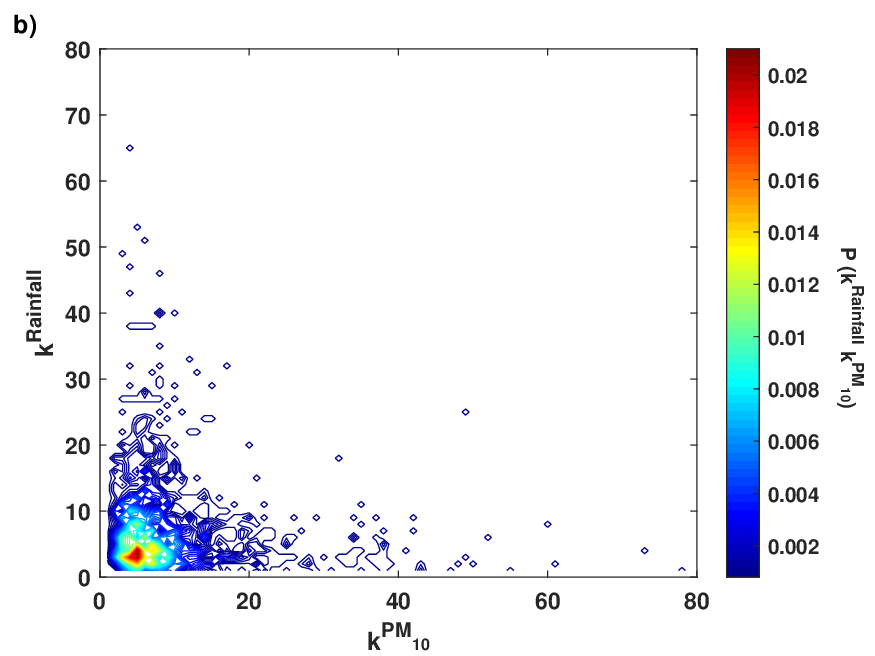}
\includegraphics[scale=0.55]{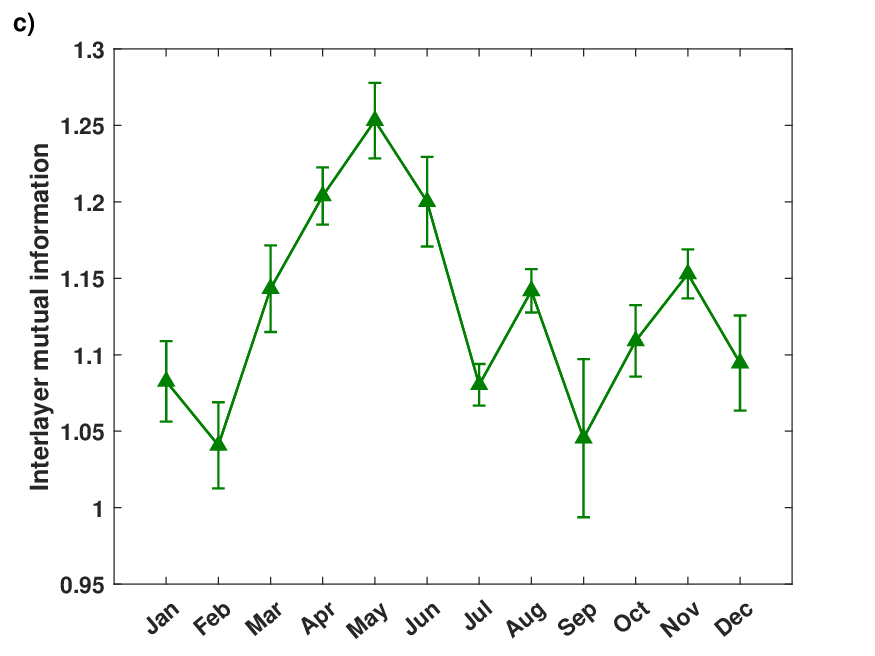}
\caption{\label{resmontinter} Illustration of the joint probability distribution of the degrees of both layers for (a) the low dust season (October to April) and (b) the high dust season (May to September) over an 11-y period. Each isoline shows the probability that the degree is precisely $k^{PM_{10}}$ and $k^{Rainfall}$ at the same time node in the VG frame; (c) monthly interlayer mutual information values ($I_{PM_{10}Rainfall}$)  and their standard deviations computed over an 11-y period.}
\end{figure}

	Figure \ref{resmontinter}(a) and \ref{resmontinter}(b) illustrate the quantity $P(k^{[PM_{10}]},k^{[Rainfall]})$ for the low dust season (October to April) and the high dust season (May to September) for the 11-y period. In these Figures, the colours indicate the probability that a node in the MVG has a degree equal to $k^{PM_{10}}$ and $k^{Rainfall}$ in the layers corresponding to $PM_{10}$ and rainfall VG, respectively. Overall, the most likely combinations of $k$ values were those below a value of 20. For higher degrees ($k > 60$), $P(k^{[PM_{10}]},k^{[Rainfall]})$ becomes less significant. The probability asymptotically approaches both the $X$ and $Y$ axes. According to \cite{carmona2020}, as the degree increases, the probability of finding $k^{PM_{10}}$ and $k^{Rainfall}$ with close values decreases exponentially. This demonstrates alternation between the hubs of the two time series. Due to the wet scavenging phenomenon, high daily values of $PM_{10}$ and rainfall are less likely to occur on the same day. Figure \ref{resmontinter}(a-b) shows a difference in behaviour between both seasons. A concentration of probability is more pronounced in the high dust season (Figure \ref{resmontinter}(b)), and higher values in a low-degree area add red to the plot. In addition, the overall shape of the plot shrinks for the same period and has shorter tails. These results are consistent with those obtained for $\omega$. The impact of rainfall on $PM_{10}$ concentrations in the atmosphere more loaded with dust from May to September is more efficient because of a more significant wet scavenging phenomenon \citep{tiwari2012, yoo2014}.
	
	Figure \ref{resmontinter}(c) shows the monthly $I_{PM_{10},Rainfall}$ computed over the 11-y period. The interlayer mutual information, which provides an idea of the typical amount of information flow in the system \citep{lacasa2015}, is directly related to the joint probability and also measures the correlation between degrees in the system. Thus, the interlayer mutual information may indicate the degree of correlation among the distributions and, hence, the behaviour of the two series. Because $I_{PM_{10},Rainfall} > 1$, $k^{PM_{10}}$ and $k^{Rainfall}$ always have higher correlations in May  ($I_{PM_{10},Rainfall}= 1.25$), August ($I_{PM_{10},Rainfall}= 1.14$) and November ($I_{PM_{10},Rainfall}= 1.15$). A study in the Caribbean basin by \cite{gouirand2020} showed that the averages transition dates from winter to summer and from summer to winter occurred on average 13 May ($\pm$ 9 days) and 26 October ($\pm$ 12 days), respectively. These transition periods correspond to comparatively high  $I_{PM_{10},Rainfall}$ values. Due to the standard deviations, the winter to summer transition always occurs in May, whereas the summer to winter transition can occur in October or November. This could explain why the May peak was much larger. In addition, May corresponds to the beginning of the high dust season \citep{plocoste2020}, whereas November corresponds to the end of the rainy season \citep{bertin2013}. Therefore, these periods feature strong inter-layer correlations between $k^{PM_{10}}$ and $k^{Rainfall}$.

\section{Conclusion}
\label{conclusion}

	In conclusion, our results clearly highlight the efficiency of multilayer complex networks for tracking the correlations between particulate matter ($PM_{10}$) and rainfall time series. The aim of this study was to investigate the wet scavenging phenomenon of $PM_{10}$ by rainfall in the Caribbean area using MVGs. We highlighted the fractal nature of both time series and found that the highest degrees (hubs) are related to the highest values in the VG frame. The relationship between the values and degrees of $PM_{10}$ is less homogeneous than that of rainfall due to annual intermittency. The monthly degree centrality analysis indicated the seasonality of both time series. On the 1-d scale, the average edge overlap ($\omega$) monthly analysis highlighted the wet scavenging process of $PM_{10}$ by rainfall throughout the year. However, this process seems to be more significant during the last six months of the year, when the high dust and rainy seasons are juxtaposed. The joint probability results between the $PM_{10}$ and rainfall nodes according to African dust seasonality confirmed the trend observed from the $\omega$ values. The atmosphere is loaded with dust during the high dust season, and rainfall helps restore the $PM_{10}$ atmospheric balance. Thus, the overall coherence in the multivariate time series was higher from July to December. The interlayer mutual information ($I_{PM_{10}Rainfall}$) monthly analysis showed a correlation between $PM_{10}$ and rainfall structures throughout the year. $I_{PM_{10}Rainfall}$ values were higher during the transition periods between winter and summer (and vice versa) in the Caribbean Basin. We assume that the transition periods allow the homogenisation of the multivariate time series before the usual trend is resumed. To better quantify the impact of the wet scavenging process on $PM_{10}$, a future analysis of rainwater chemistry (organic and elemental carbon) related to rainfall intensity will be conducted.

\section*{Acknowledgements}
The authors are very grateful to the anonymous reviewers for their valuable comments and constructive suggestions, which helped us to improve the quality of the paper substantially. The authors would like to thank Guadeloupe air quality network (Gwad'Air) and the French Weather Office (M\'et\'eo France Guadeloupe) for providing air quality and meteorological data. A special thanks to Mr. France-Nor Brute (Data Scientist) for data assistance. Group TEP-957 authors gratefully acknowledge support of the Research Program of the University of Cordoba (2021), Spain.

\section*{Disclosure statement}

No potential conflict of interest was reported by the authors.

\section*{Funding}

The authors declare that they have not received any fund for the present paper. The paper is the sole work of the authors and is not a part/product of any project. 

\clearpage

\section*{References}
\bibliographystyle{model4-names}
\biboptions{authoryear}

\bibliography{Washout}

\end{document}